\documentclass[11pt,a4paper]{article}
\usepackage[hyperref]{acl2021}
\usepackage{times}
\usepackage{latexsym}

\usepackage{microtype}
\usepackage{graphicx}
\usepackage{booktabs}  
\usepackage{multirow}
\usepackage{subfigure}
\usepackage{verbatim}
\usepackage{CJKutf8} 
\usepackage{flushend}
\usepackage{makecell}
\usepackage{amsmath}
\usepackage{amssymb}
\aclfinalcopy 

\title{Is News Recommendation a Sequential Recommendation Task?}

\author{Chuhan Wu$^\dagger$~~~~Fangzhao Wu$^\ddagger$~~~~Tao Qi$^\dagger$~~~~Yongfeng Huang$^\dagger$\\
    $^\dagger$Department of Electronic Engineering \& BNRist, Tsinghua University, Beijing 100084, China  \\
     $^\ddagger$Microsoft Research Asia, Beijing 100080, China\\
  {\tt\{wuchuhan15, wufangzhao, taoqi.qt\}@gmail.com} \\ {\tt yfhuang@tsinghua.edu.cn}
  }
\date{}

\begin{document}
\maketitle

\begin{abstract}

News recommendation is often modeled as a sequential recommendation task, which assumes that there are rich short-term dependencies over historical clicked news.
However, in news recommendation scenarios users usually have strong preferences on the temporal diversity of news information and may not tend to click similar news successively, which is very different from many  sequential recommendation scenarios such as e-commerce recommendation.
In this paper, we study whether news recommendation can be regarded as a standard sequential recommendation problem.
Through extensive experiments on two real-world datasets, we find that modeling news recommendation as a sequential recommendation problem is suboptimal.
To handle this challenge, we further propose a temporal diversity-aware news recommendation method that can promote candidate news that are diverse from recently clicked news, which can help predict future clicks more accurately.
Experiments show that our approach can consistently improve various news recommendation methods.

\end{abstract}

\section{Introduction}

News recommendation is important for improving users' online news reading experience~\cite{wu2020mind}.
Many existing news recommendation methods model the new recommendation task as a sequential recommendation problem~\cite{wu2019nrms,gabriel2019contextual,wu2021fairness,qi2021pprec}.
For example, \citet{okura2017embedding} use a GRU network to model user interest from the sequence of clicked news,  and rank candidate news based on their relevance to user interest.
\citet{zhu2019dan} use a combination of LSTM network and directional self-attention network to learn user interest representations from clicked news sequence, and further match it with candidate news.
\citet{wu2020sentirec} use a Transformer to model clicked news sequence to learn user interest representation for interest matching. 
A core assumption of these methods is that there are rich short-term dependencies over historical behaviors, and future behaviors are also likely to be relevant to  recent past behaviors~\cite{hidasi2016session}.
Although this assumption is widely used by many sequential recommendation scenarios like e-commerce recommendation~\cite{chen2018sequential} and movie recommendation~\cite{kang2018self}, we find it may not be valid in the news recommendation scenario due to users' preference on the temporal diversity (i.e., novelty) of news information~\cite{garcin2013personalized}.
For example, in the MIND~\cite{wu2020mind} news recommendation dataset, only 7.2\% adjacently clicked news are in the same topic category (the ratio is 7.9\% for random clicks).
In addition, only 0.04\% adjacently clicked news mention the same entities, while 0.11\% random pairs of clicked news share at least one same entity.\footnote{We observe similar phenomena in our production news recommendation dataset.}
These results show that adjacently clicked news tend to be diverse rather than similar, which contradicts the  basic assumption of sequential recommendation.

In this paper, we study whether news recommendation is suitable to be modeled as a sequential recommendation task.
Through extensive experiments on two real-world news recommendation datasets, we find that many mainstream sequence modeling techniques used by existing news recommendation methods are not optimal, which indicates that modeling news recommendation as sequential recommendation is suboptimal.
To solve this problem, we propose a temporal diversity-aware news recommendation method named \textit{TempRec}, which encourages to recommend candidate news that are diverse from recently clicked news.
More specifically, we use a shared order-agnostic Transformer to learn a global interest embedding from all  historical news clicks  and learn a recent interest embedding from several latest clicks.
The click score is computed by a learnable weighted summation of the positive relevance between candidate news and global interest embedding as well as the negative relevance between candidate news and recent interest embedding.
Experimental results demonstrate that \textit{TempRec} can outperform many existing news recommendation methods.

\begin{figure}[!t]
  \centering 
      \includegraphics[width=0.99\linewidth]{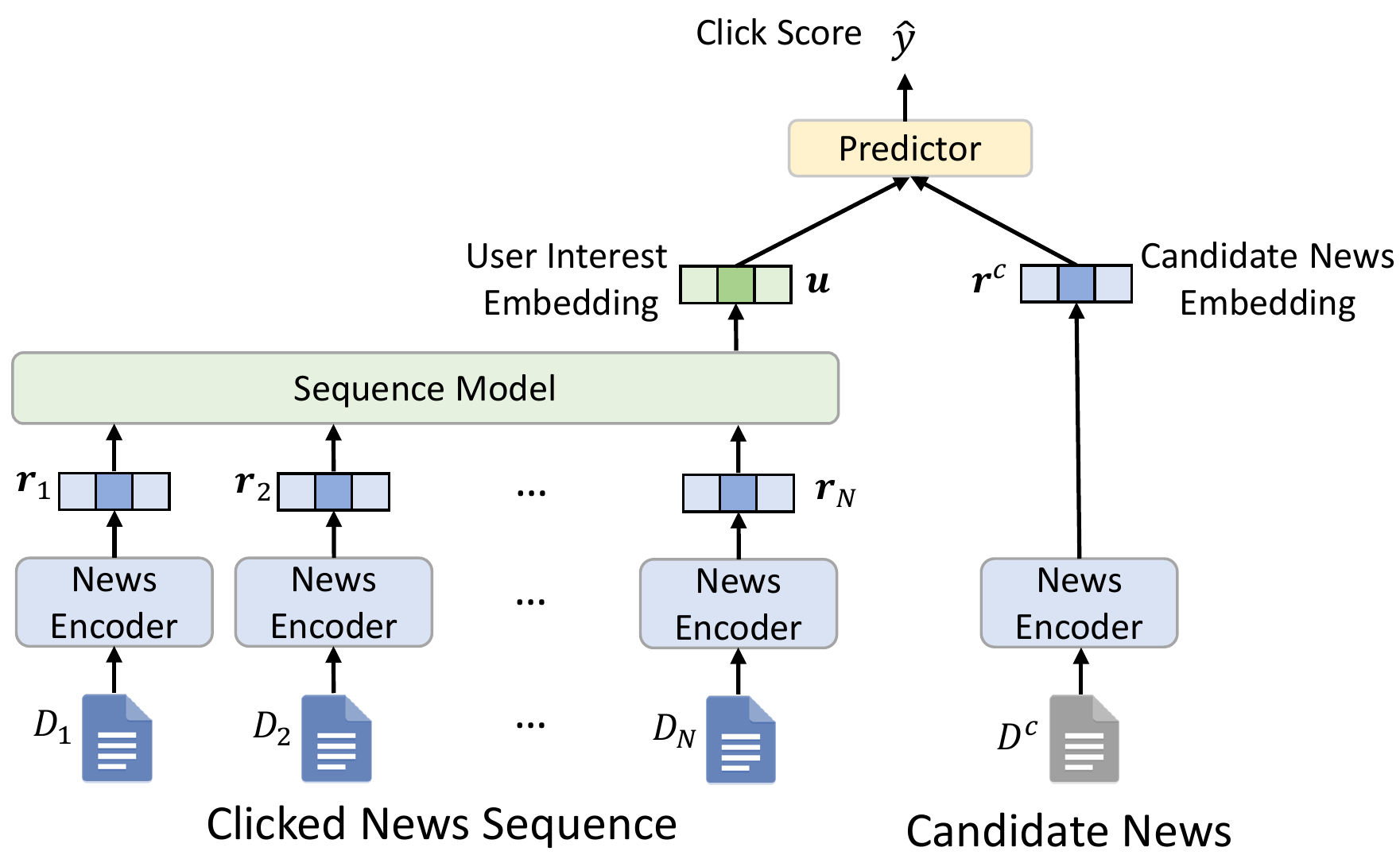}
  \caption{A general framework of sequential news recommendation.}\label{fig.model}
\end{figure}

\section{Sequential News Recommendation}

In this section, we introduce the general framework of sequential news recommendation, which is shown in Fig.~\ref{fig.model}.
Assume a user has $N$ historical clicked news, which forms a news sequence $[D_1, D_2, ..., D_N]$.
The candidate news is denoted as $D^c$.
The framework first uses a news encoder to learn news representations from news texts, which can be implemented by various models such as autoencoder~\cite{okura2017embedding}, CNN~\cite{wu2019} and self-attention network~\cite{wu2019nrms}.
We denote the clicked news representation sequence learned by news encoders as $[\mathbf{r}_1, \mathbf{r}_2, ..., \mathbf{r}_N]$ and candidate news representation as $\mathbf{r}^c$.
The framework then uses a sequence model to learn a user interest embedding $\mathbf{u}$ from the clicked news representation sequence.
It can be typically implemented by sequence modeling techniques such as LSTM~\cite{zhu2019dan}, GRU~\cite{an2019neural} and self-attention~\cite{wu2019nrms}. 
The framework finally predicts the click score $\hat{y}$ for personalized ranking based on the relevance between user interest embedding and candidate news embedding, which is usually evaluated by the inner product between them.
For model training, negative sampling techniques~\cite{wu2019npa} are used for constructing training samples from click logs, and crossentropy is used as the loss function for model optimization.
\section{Experiments on Sequential News Recommendation}

\begin{table}[!t]
\centering
\begin{tabular}{lcc}
\Xhline{1.5pt}
                     & MIND       & News       \\ \hline
\#users              & 1,000,000  & 874,891    \\
\#news               & 161,013    & 1,322,973  \\
\#impressiom         & 15,777,377 & 1,000,000  \\
\#click behaviors    & 24,155,470 & 41,976,699 \\
avg. title len.      & 11.52      & 12.62      \\
avg. click seq. len. & 36.72      & 40.18      \\ \Xhline{1.5pt}
\end{tabular}
\caption{Statistics of \textit{MIND} and \textit{News} datasets.}\label{dataset}
\end{table}

\begin{table*}[t]
\resizebox{1\linewidth}{!}{ 
\begin{tabular}{lcccccccc}
\Xhline{1.5pt}
\multicolumn{1}{c}{\multirow{2}{*}{\textbf{Model}}} & \multicolumn{4}{c}{\textbf{MIND}} & \multicolumn{4}{c}{\textbf{News}} \\ \cline{2-9} 
\multicolumn{1}{c}{}                                & AUC    & MRR   & nDCG@5 & nDCG@10 & AUC    & MRR   & nDCG@5 & nDCG@10 \\ \hline
LSTUR                                               & 68.43  & 33.66 & 36.61  & 42.27   & 64.45  & 36.73 & 39.98  & 45.85   \\
LSTUR (inverse)                                     & 68.49  & 33.68 & 36.65  & 42.31   & 64.43  & 36.72 & 39.94  & 45.82   \\
LSTUR (random)                                      & 68.76  & 33.94 & 36.89  & 42.55   & 64.67  & 36.98 & 40.21  & 46.03   \\ \hline
DAN                                                 & 67.96  & 33.22 & 36.18  & 41.95   & 63.92  & 36.30 & 39.61  & 45.46   \\
DAN (inverse)                                       & 67.99  & 33.26 & 36.21  & 41.97   & 63.95  & 36.32 & 39.64  & 45.47   \\
DAN (random)                                        & 68.30  & 33.54 & 36.47  & 42.18   & 64.29  & 36.54 & 39.87  & 45.69   \\ \hline
NRMS                                                & 68.22  & 33.46 & 36.49  & 42.15   & 64.22  & 36.59 & 39.87  & 45.64   \\
NRMS+PE                                             & 68.13  & 33.38 & 36.40  & 42.07   & 64.27  & 36.66 & 39.92  & 45.69   \\
NRMS+PE (inverse)                                   & 68.15  & 33.40 & 36.39  & 42.10   & 64.23  & 36.64 & 39.89  & 45.65   \\
NRMS+PE (random)                                    & 68.20  & 33.43 & 36.46  & 42.17   & 64.25  & 36.66 & 39.91  & 45.67   \\
NRMS+CM                                             & 67.67  & 33.04 & 35.92  & 41.74   & 63.87  & 36.24 & 39.55  & 45.36   \\
NRMS+CM (inverse)                                   & 67.71  & 33.09 & 35.95  & 41.78   & 63.91  & 36.26 & 39.55  & 45.40   \\
NRMS+CM (random)                                    & 68.06  & 33.27 & 36.31  & 41.92   & 64.11  & 36.51 & 39.77  & 45.52   \\ \Xhline{1.5pt}
\end{tabular}
}

\caption{Performance of different methods. PE denotes position embedding and CM denotes casual mask.} \label{table.performance} 
\end{table*}

\subsection{Datasets and Experimental Settings}

We conduct extensive experiments on two real-world news recommendation datasets to compare different sequential news recommendation methods.
The first dataset is MIND~\cite{wu2020mind}\footnote{https://msnews.github.io/}, which is a large-scale benchmark news recommendation dataset.
The second dataset (denoted as \textit{News}) is constructed by ourselves based on 1 million news impression logs from 10/17/2020 to 01/29/2021 on a commercial news platform.
The statistics of the two datasets are shown in Table~\ref{dataset}.
On both datasets, the logs in the last week are used for training, logs in the day before the last week are used for validation, and the rest are for training.

In our experiments, we use Glove~\cite{pennington2014glove} word embeddings for initialization.
The hidden dimension of all models is 400.
We use Adam~\cite{kingma2014adam} (lr=1e-4) as the optimizer.
Following~\cite{wu2020mind}, we use AUC, MRR, nDCG@5 and nDCG@10 as the performance metrics.
We report the average results of 5 independent  experiments.

\subsection{Performance Comparison}

We compare three widely used sequential news recommendation baselines, including: (1) LSTUR~\cite{an2019neural}, which uses GRU to model short-term user interest and uses user ID embedding to model long-term user interest;
(2) DAN~\cite{zhu2019dan}, which uses a combination of LSTM and casual self-attention network to model user interest;
(3) NRMS~\cite{wu2019neuralnrms}, which uses multi-head self-attention to model user interest.
Note that the original NRMS model does not incorporate position information.
For NRMS, we also compare its two order-aware variants that incorporate learnable position embeddings or casual self-attention mask~\cite{kang2018self}.
In addition, to further explore whether sequential information can benefit news recommendation, we compare two variants of all methods that use inverse or randomly shuffled clicked news sequence.
The performance on two datasets is shown in Table~\ref{table.performance}.
We have several interesting observations.
First, compared with the order-agnostic NRMS model, incorporating position embedding does not yield performance gain, and casual self-attention is inferior to the bidirectional self-attention.
This indicates that positional information may not be very important for understanding user interest in news recommendation and it is better to fully model past and future information in the clicked news sequence.
Second, we find that using inverse sequence does not lead to a notable performance difference.
Since recurrent models usually pay more attention to  latest steps, it implies that in news recommendation recent news clicks are not more informative than earlier clicks in predicting future clicks, which is probably due to the effects of users' temporal diversity preferences and their stable long-term interest.
Moreover, it is very interesting that randomly shuffle the click sequence can even slightly improve the performance of order-sensitive models.
This may be because the model can better capture global user interest from the shuffled sequence to help predict future clicks more accurately.
These results demonstrate that news recommendation may not be suitable to be modeled as a sequential recommendation problem because neither order information nor short-term dependencies plays an important role in news recommendation.

\begin{figure}[!t]
  \centering 
      \includegraphics[width=0.99\linewidth]{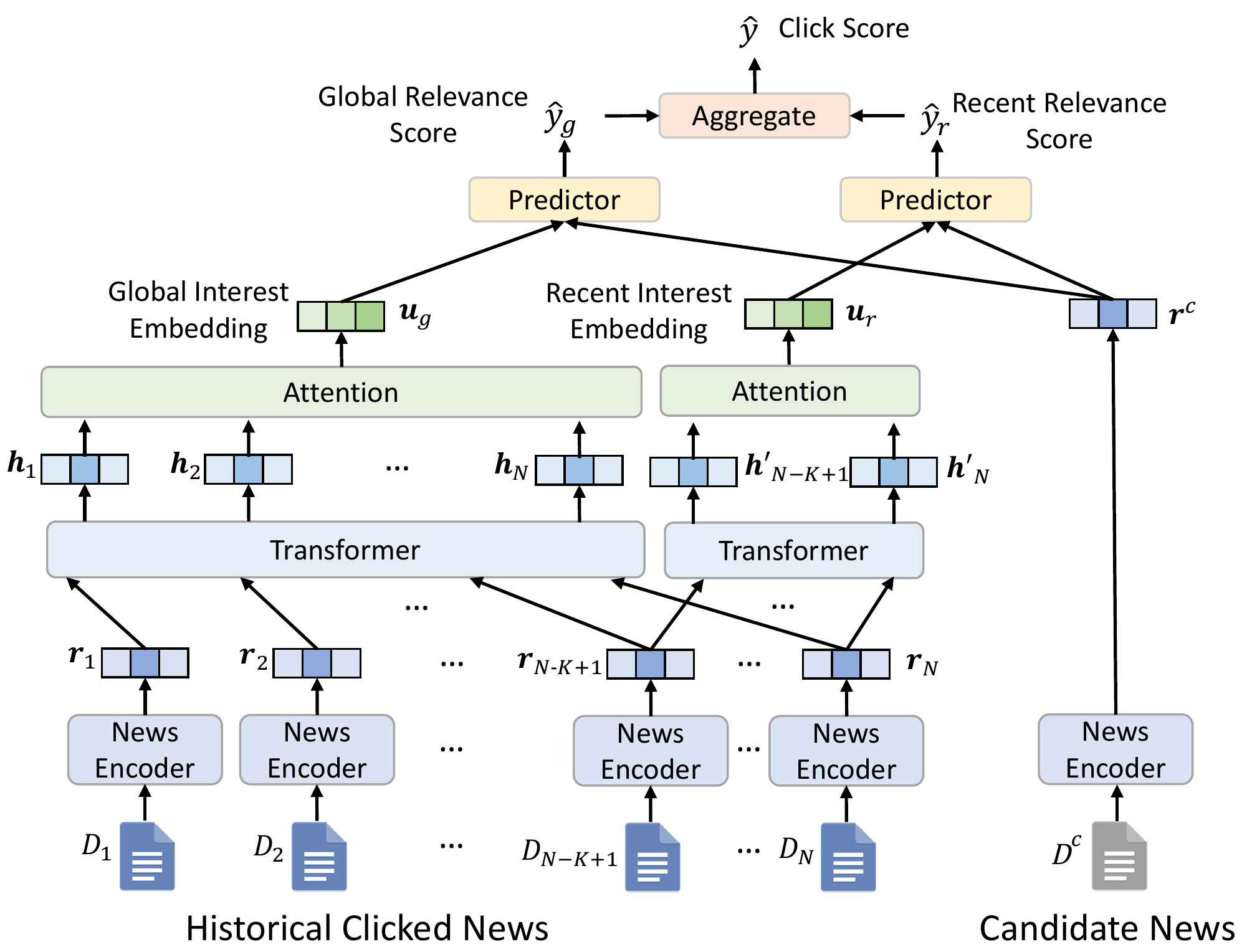}
  \caption{The architecture of \textit{TempRec}.}\label{fig.model2}
\end{figure}

\section{Temporal Diversity-aware News Recommendation}

According to the experimental results and analysis in the previous section, it is not optimal to use sequential models to process clicked news sequences for news recommendation.
To handle this challenge, we propose a novel temporal diversity-aware news recommendation method named \textit{TempRec}, which can consider the temporal diversity nature of news recommendation to make more accurate recommendation results.
The architecture of \textit{TempRec} is shown in Fig.~\ref{fig.model2}.
There are two order-agnostic Transformers in \textit{TempRec}.\footnote{We share their parameters to control the model size.}
One of them is used to process the entire clicked news sequence, which aims to capture global user interest.
The other one is used to process the latest $K$ news clicks, which aims to capture recent user interest.
We denote the hidden news representation sequences learned by them as $\mathbf{H}=[\mathbf{h}_1, \mathbf{h}_2, ..., \mathbf{h}_N]$ and $\mathbf{H}'=[\mathbf{h}'_{N-K+1}, ..., \mathbf{h}'_N]$, respectively.
We then use two attention networks to select important news clicks in the two sequences to learn a global interest embedding $\mathbf{u}_g$ from $\mathbf{H}$ and learn a recent interest embedding $\mathbf{u}_r$ from $\mathbf{H}'$.
Since future clicked news may tend to be diverse from recently clicked news, we propose a temporal diversity-aware click prediction method to help predict news clicks more accurately.
More specifically, we compute a global relevance score $\hat{y}_g$ based on the relevance between candidate news embedding and global interest embedding (i.e., $\hat{y}_g=\mathbf{u}_g\cdot \mathbf{r}^c$), which indicates whether candidate news matches the overall user interest.
In addition, we compute a recent relevance score $\hat{y}_r$ via $\hat{y}_r= \mathbf{u}_r \cdot \mathbf{r}^c$, which indicates the relevance between candidate news and recent click behaviors.
The unified click score $\hat{y}$ is a linear combination of two scores, which is formulated as $\hat{y}=\hat{y}_g-max(w,0)\hat{y}_r$, where $w$ is a  learnable parameter.
In this way, the model can encourage to recommend news that are diverse from recently clicked news, which can better satisfy users' need on the temporal diversity of news information.

\section{Experiments on TempRec}

\begin{figure}[!t]
  \centering
  \subfigure[\textit{MIND}.]{
    \includegraphics[width=0.48\textwidth]{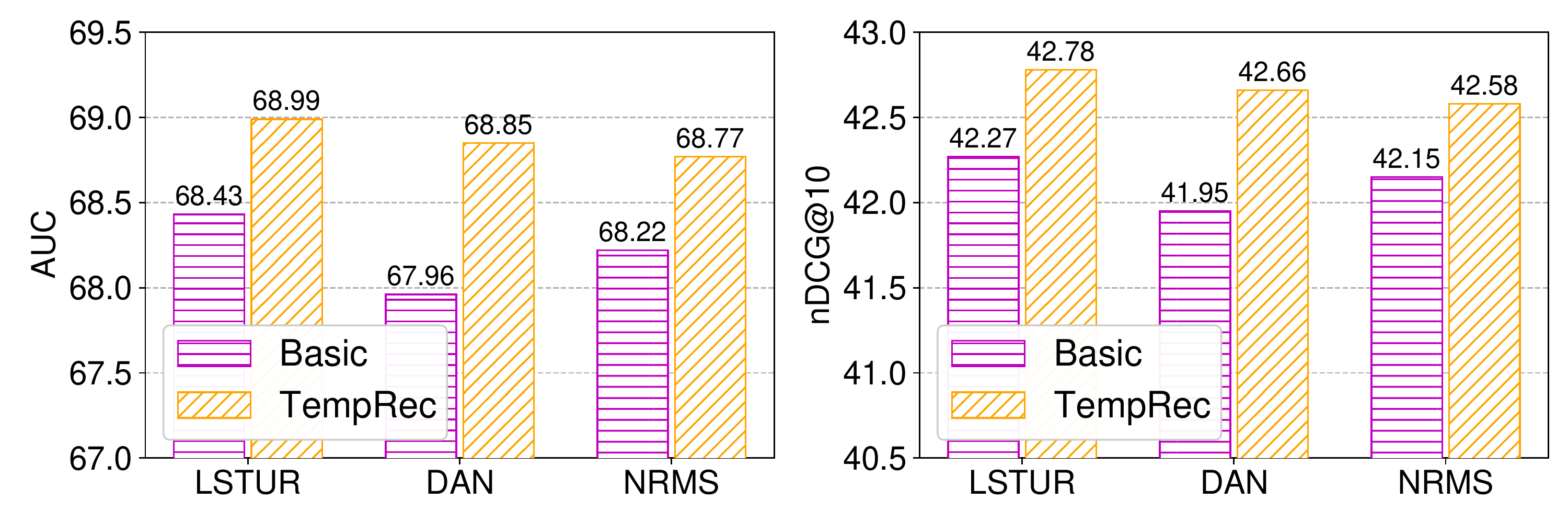}
    }  
    \subfigure[\textit{News}.]{
    \includegraphics[width=0.48\textwidth]{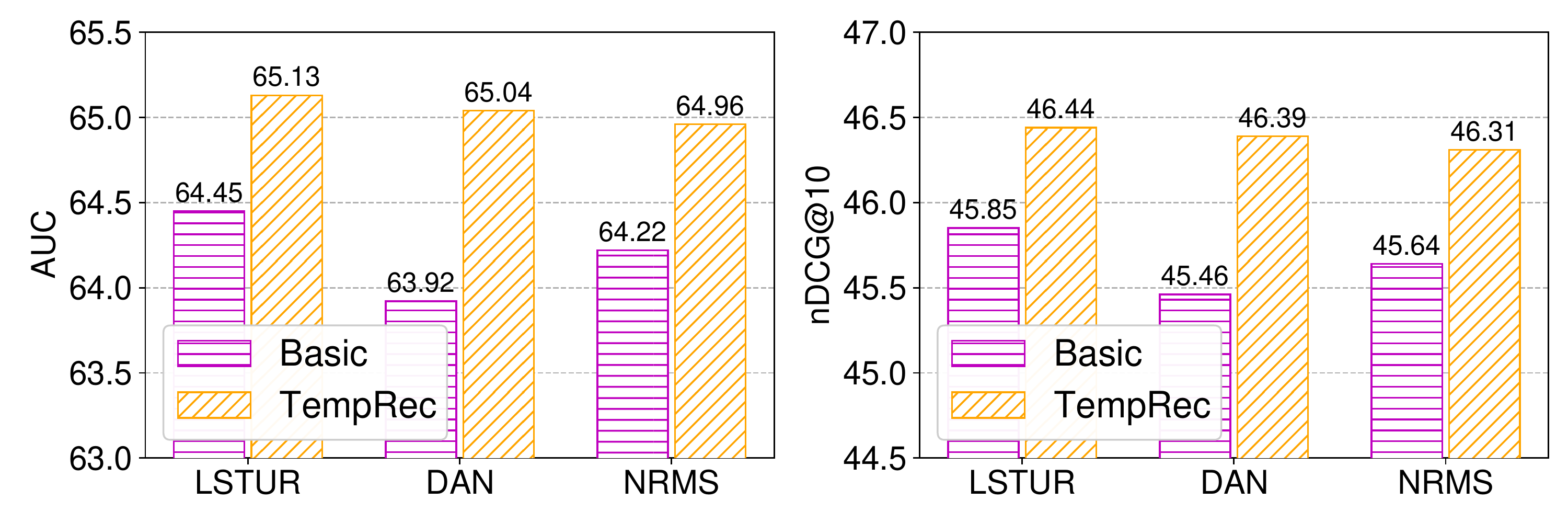}
    }
  \caption{Effectiveness of \textit{TempRec}.}\label{fig.ab}

\end{figure}

We further conduct experiments to verify the effectiveness of \textit{TempRec}.
Since \textit{TempRec} is a general framework and is compatible with many different news encoders, we compare the performance of \textit{TempRec} by using the same news encoders with \textit{LSTUR}, \textit{DAN} and \textit{NRMS}.
The results on the two datasets are shown in Fig.~\ref{fig.ab}.
We find that \textit{TempRec} can consistently improve the performance of different methods, and further t-test results show the improvements are significant ($p<0.05$).
These results show that considering the temporal diversity characteristics of users' news click behaviors can help predict future news clicks more accurately.
In addition, we find the parameter $w$ is 0.075 and 0.083 on the \textit{MIND} and \textit{News} datasets, respectively.
It shows that very recent clicks have slight negative relevance to future clicks.

\begin{figure}[!t]
  \centering
  \subfigure[\textit{MIND}.]{
    \includegraphics[width=0.22\textwidth]{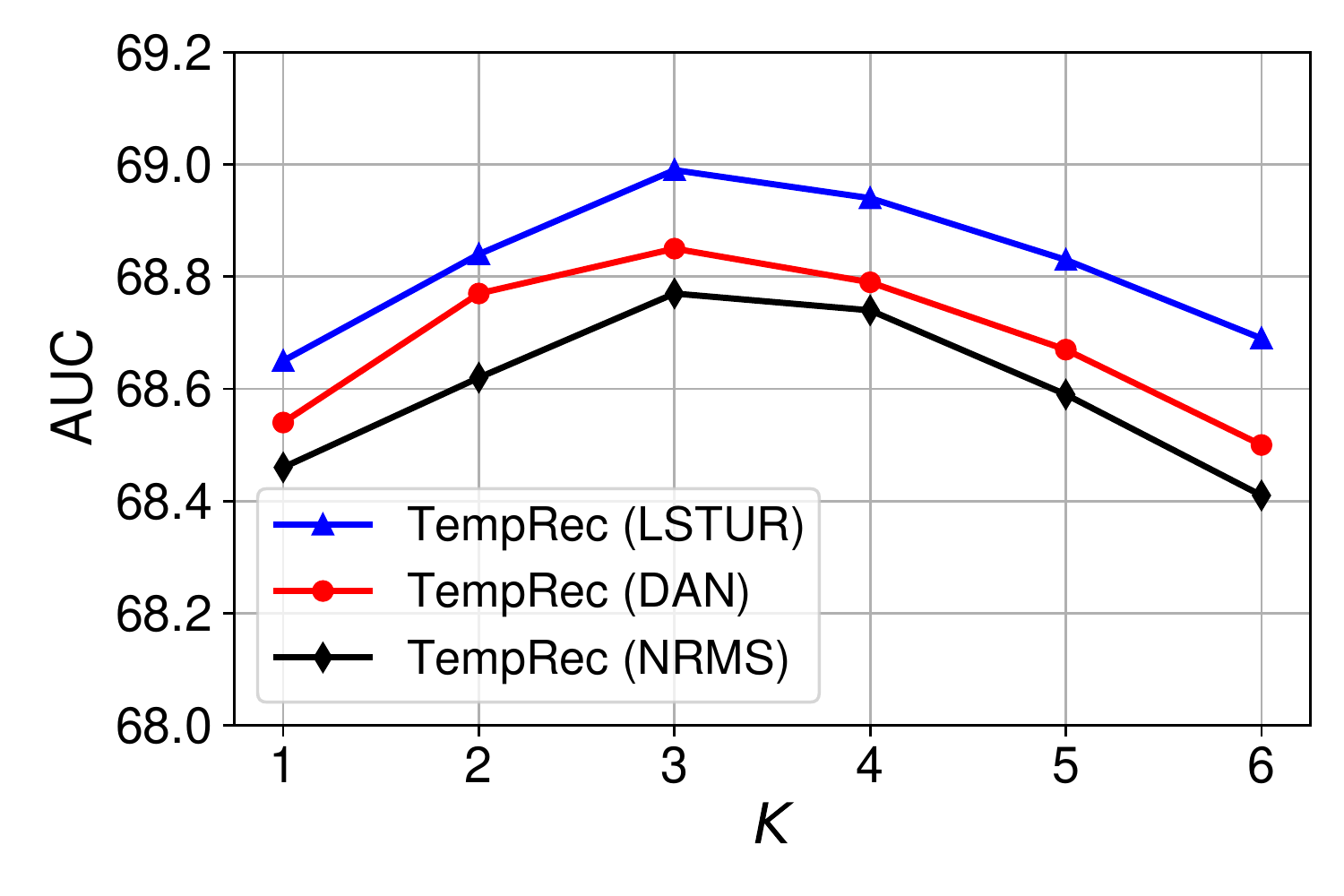}
    }  
    \subfigure[\textit{News}.]{
    \includegraphics[width=0.22\textwidth]{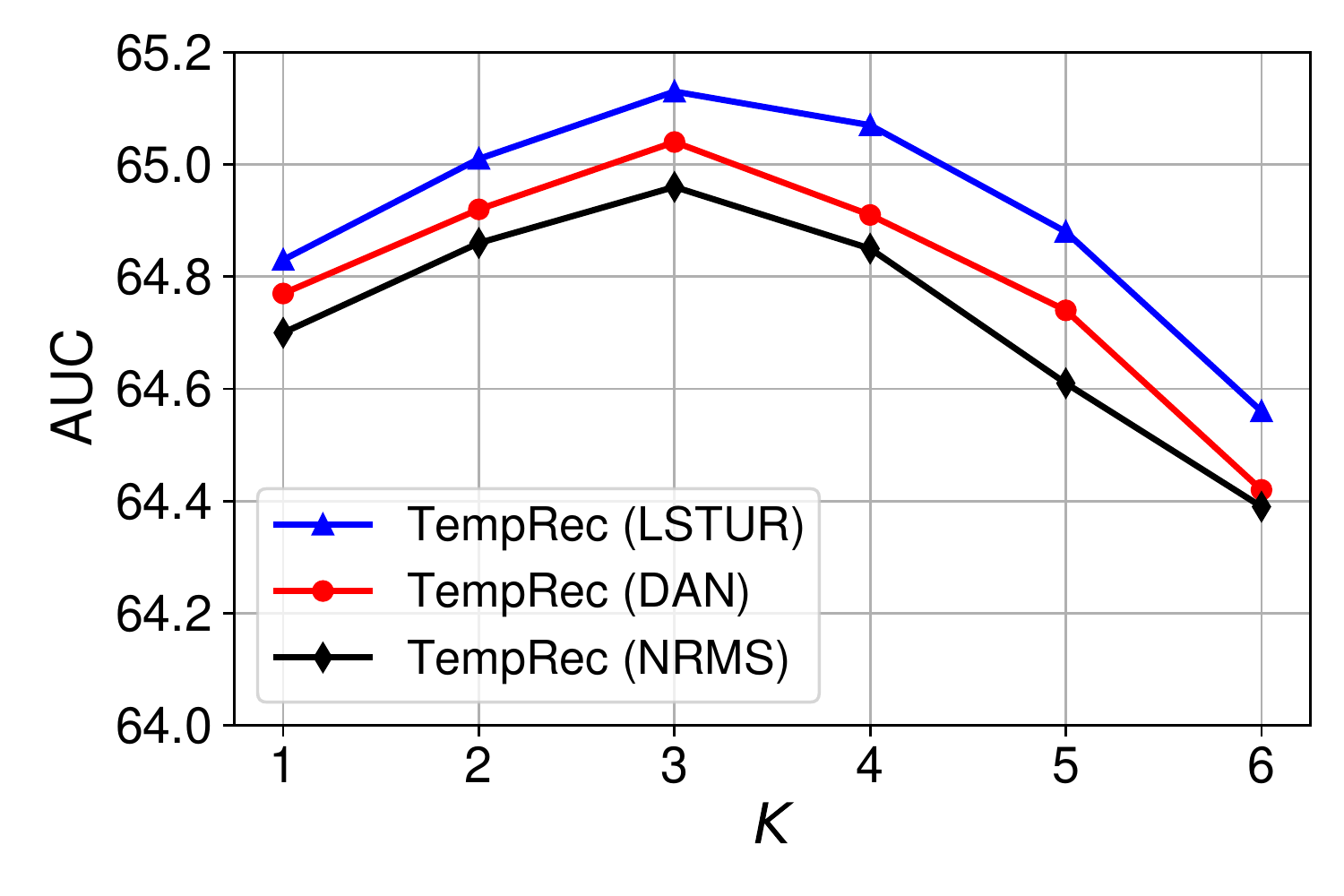}
    }
  \caption{Influence of the hyperparameter $K$.}\label{fig.hyper}

\end{figure}

We then study the influence of the hyperparameter $K$ on model performance, as shown in Fig.~\ref{fig.hyper}.
We find that $K=3$ is an appropriate choice on both datasets.
This is probably because the temporal diversity between future and past clicks cannot be fully modeled when $K$ is too small, while relevant clicks may not be effectively matched when $K$ is too large.
Thus, we set $K=3$ that yields the best performance.

\section{Conclusion}\label{sec:Conclusion}

In this paper, we study an interesting and important problem, i.e., whether news recommendation is suitable to be modeled as sequential recommendation.
Through extensive experiments on two real-world datasets, we find that many mainstream sequence models used by existing methods are suboptimal for news recommendation, and news recommendation does not satisfy the basic assumption of sequential recommendation.
To address this challenge, we propose a temporal diversity-aware news recommendation method named \textit{TempRec} that can consider the temporal diversity nature of users' news click behaviors to make better recommendations.
Experimental results on the two datasets validate the effectiveness of \textit{TempRec}.


\bibliographystyle{acl_natbib}
\bibliography{acl2021}


\end{document}